\documentclass[useAMS,usenatbib]{mn2e}

\usepackage{color}
\usepackage{graphicx}
\usepackage{graphics}
\usepackage{rotating}
\usepackage{amsmath}        
\usepackage{amssymb}        
\usepackage{subfigure}        
\usepackage{epsfig}        

\newcommand{\degree}{$^{\circ}$}

\newcommand{\arcsecond}{$^{\prime\prime}$}
\newcommand{\perbeam}{\,beam$^{-1}$}

\title[Resolved au-scale radio jets in Circinus X-1]{The first resolved imaging of milliarcsecond-scale jets in Circinus X-1}

\author[J.C.A.~Miller-Jones et al.]
 {J.C.A.~Miller-Jones$^{1}$\thanks{email: james.miller-jones@curtin.edu.au},
 A.~Moin$^{1}$, S.J.~Tingay$^1$, C.~Reynolds$^1$, C.J.~Phillips$^{2}$, \and A.K.~Tzioumis$^{2}$, R.P.~Fender$^{3}$, J.N.~McCallum$^{4}$, G.D.~Nicolson$^{5}$, and V.~Tudose$^{6}$\\
$^1$International Centre for Radio Astronomy Research - Curtin University, GPO Box U1987, Perth, WA 6845, Australia\\
$^{2}$CSIRO Astronomy and Space Science, PO Box 76, Epping, NSW, Australia \\
$^{3}$School of Physics and Astronomy, University of Southampton, High Field SO17 IBJ, England\\
$^{4}$School of Mathematics and Physics, University of Tasmania, Private Bag 21, Hobart, Tasmania 7001, Australia\\
$^{5}$ Hartebeesthoek Radio Astronomy Observatory, PO Box 443, Krugersdorp 1740, South Africa \\
$^{6}$Netherlands Institute for Radio Astronomy, Oude Hoogeveensedijk 4, 7991 PD Dwingeloo, the Netherlands
}
\begin{document}

\date{Accepted 2011 October 13.  Received 2011 October 13; in original form 2011 June 9}

\pagerange{\pageref{firstpage}--\pageref{lastpage}} \pubyear{2011}

\maketitle

\label{firstpage}

\begin{abstract}
We present the first resolved imaging of the milliarcsecond-scale jets in the neutron star X-ray binary Circinus X-1, made using the Australian Long Baseline Array.  The angular extent of the resolved jets is $\sim20$\,milliarcseconds, corresponding to a physical scale of $\sim150$\,au at the assumed distance of 7.8\,kpc.  The jet position angle is relatively consistent with previous arcsecond-scale imaging with the Australia Telescope Compact Array.  The radio emission is symmetric about the peak, and is unresolved along the minor axis, constraining the opening angle to be $<20$\degree.  We observe evidence for outward motion of the components between the two halves of the observation.  Constraints on the proper motion of the radio-emitting components suggest that they are only mildly relativistic, although we cannot definitively rule out the presence of the unseen, ultra-relativistic ($\Gamma>15$) flow previously inferred to exist in this system.
\end{abstract}


\section{Introduction}

Circinus X-1 (Cir X-1) is one of the few confirmed Galactic neutron star X-ray binaries with resolved radio jets, which have been studied from sub-arcsecond \citep{Cal11} through arcsecond \citep{Fen98,Tud08} and out to arcminute scales \citep{Ste93,Tud06}, where they have inflated a synchrotron-emitting radio nebula.  Cir X-1 is also the only confirmed neutron star system with resolved X-ray jets, oriented along the same position angle as the radio jets \citep{Hei07,Sol09}.  The system comprises a neutron star primary \citep{Lin10} accreting from a less-evolved companion whose nature is relatively poorly constrained.  The binary has an eccentric 16.6 day orbit \citep{Kal76}, with enhanced accretion close to periastron \citep{Mur80,Hay80} leading to flaring events in the radio, infrared and X-ray bands.  The distance to Cir X-1 is not well-constrained.  From the peak luminosity of its Type I X-ray bursts, \citet{Jon04} estimated a distance in the range 7.8--10.5\,kpc, in agreement with the earlier H\textsc{i} absorption distance of $>8$\,kpc determined by \citet{Gos77}.  However, \citet{Iar05} claimed a lower value of 4.1\,kpc from model-fitting of X-ray spectra that suggested a low hydrogen column towards the source.

The orientation of the system remains controversial.  The presence of X-ray dips \citep{Shi99} and the spectral changes on egress \citep{Bra96}, as well as the P-Cygni profiles of detected X-ray lines \citep{Bra00}, argue for an edge-on accretion disc.  However, the extremely high Lorentz factor ($\Gamma>15$) inferred from the time delay between a flare in the radio core and subsequent downstream flaring in the jets would require the radio jets to be inclined very close to the line of sight \citep{Fen04}.  Reconciling all these constraints would require a bent jet or a severe misalignment between the jet axis and the orbital plane \citep[as discussed for other systems by][]{Mac02}.

The ultra-relativistic flow inferred by \citet{Fen04} depends on the assumption of a causal relation between flaring events in the core and those subsequently observed downstream in the jets.  However, the sparse time coverage of the existing radio observations has thus far precluded independent confirmation of this result.  With the high angular resolution available using very long baseline interferometry (VLBI), we can directly measure the proper motions of individual jet components, hence constraining the jet Lorentz factor and inclination angle.  However, compact radio emission is seen only during the radio flares immediately following periastron \citep{Moi11}, and the low surface brightness sensitivity of VLBI arrays constrains the observations to periods of strong radio flaring in the system.

From 1975--1985, the radio flares regularly reached levels in excess of 1\,Jy, allowing \citet{Pre83} to constrain the angular size of the flaring component to lie between 1.5 and 15\,mas at 2.3\,GHz.  After 1985, the source entered a more quiescent phase with flares peaking at mJy levels, and further activity was not seen again until 2006, when seven flares ranging from 0.2--1\,Jy were detected at the Hartebeesthoek Radio Astronomy Observatory \citep[HartRAO; ][]{Nic07}.  This triggered new VLBI observations of Cir X-1, and in the first Southern Hemisphere electronic VLBI experiment, \citet{Phi07} detected a single, scatter-broadened component of size $60\pm15$\,mas at 1.6\,GHz.  The source was also unresolved in the 1.4- and 1.7-GHz e-VLBI observations of \citet{Moi11}.

In 2010 May, enhanced X-ray activity was detected from Cir X-1 \citep{Nak10}.  Follow-up radio observations were carried out at HartRAO (Nicolson, priv.\ comm.) and subsequently at the Australia Telescope Compact Array \citep[ATCA;][]{Cal10}, measuring flux densities of order several hundred mJy, an order of magnitude higher than seen at the ATCA over the past 15\,y.  This prompted us to trigger new VLBI observations at a higher frequency and hence improved angular resolution, in an attempt to determine the milliarcsecond-scale morphology of the source.

\section{Observations and data reduction}
\label{sec:observations}

\begin{figure}
\centering
\includegraphics[width=0.95\columnwidth]{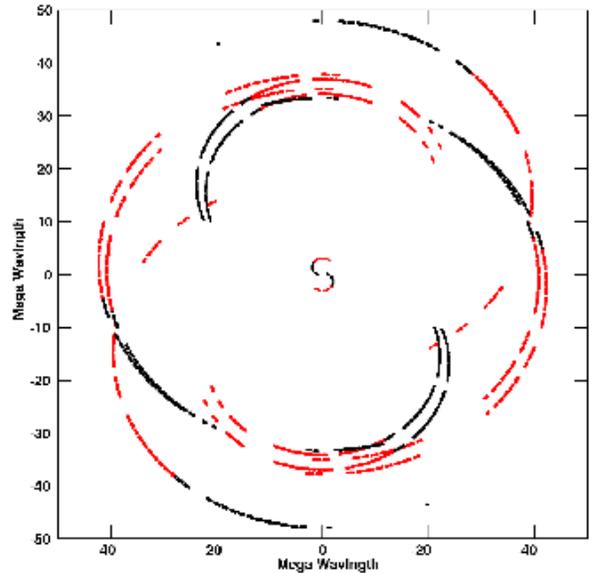}
\caption{{\it uv}-coverage of our observations.  Red points indicate the first half of the observations, and black points the second half.}
\label{fig:uvcov}
\end{figure}

We conducted a target-of-opportunity observation of Cir X-1 using the Australian Long Baseline Array (LBA) from 02:21--16:41\,UT on 2010 July 28 (orbital phase 0.046--0.082 according to the latest ephemeris from G.D.~Nicolson, priv.\ comm.).  We observed at the relatively high frequency of 8.4\,GHz, both reducing the angular broadening caused by interstellar scattering and increasing the achievable angular resolution.  The array comprised the ATCA (in its hybrid H168 configuration), Ceduna, Hobart, Mopra and Tidbinbilla telescopes.  The ATCA was used in tied-array mode, phasing together the four antennas CA01, CA02, CA03 and CA05 (maximum baseline of 192\,m).  No fringes were detected from Tidbinbilla, reducing our effective LBA array to four stations.  The resulting {\it uv}-coverage is shown in Fig.~\ref{fig:uvcov}.  The observations were conducted in dual-polarization format with 64\,MHz of bandwidth per polarization.

The data were correlated using the DiFX software correlator \citep{Del07,Del11}, and reduced using {\sc aips} \citep{Gre03}. The source amplitudes were determined using the phase calibrator source J1515-5559, assuming an approximate flux density of 1.7\,Jy from recent ATCA monitoring (with an estimated error of $\pm20$ per cent), an estimate confirmed to within the uncertainties from the ATCA data taken in VLBI mode during the LBA observations.  Since Cir X-1 was sufficiently bright, fringe fitting was performed on the target source.  Imaging and self-calibration were then carried out in {\sc Difmap} \citep*{She94}.

\section{Results}
\label{sec:results}

\begin{figure*}
\begin{center}
\mbox{\subfigure[First half of the observations (rms noise 1.51\,mJy\perbeam).  Total flux density of the jet inside the $3\sigma$ contour is $146\pm14$\,mJy.]{\epsfig{figure=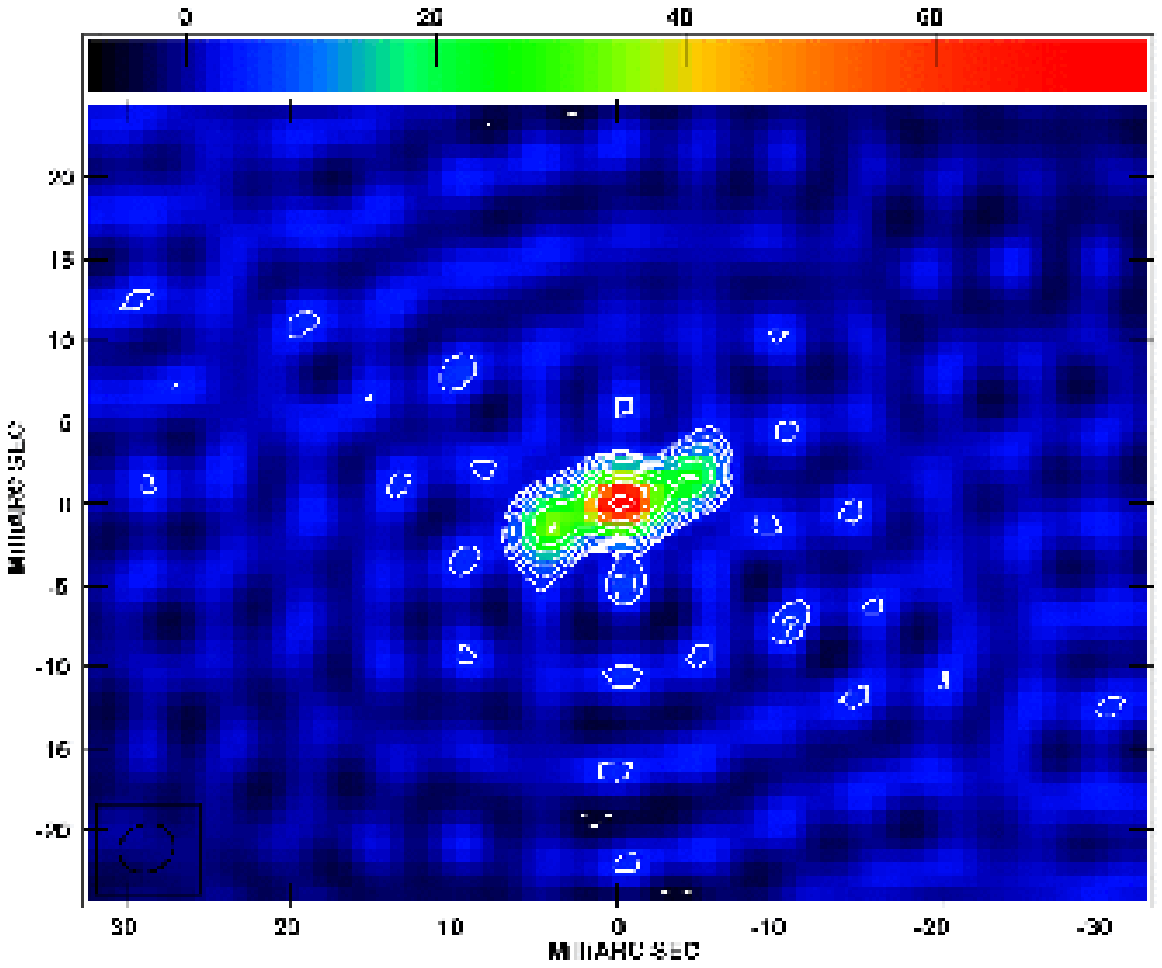,width=0.45\textwidth}}\quad\quad
  \subfigure[Second half of the observations (rms noise 0.75\,mJy\perbeam).  Total flux density of the jet inside the $3\sigma$ contour is $104\pm8$\,mJy.]{\epsfig{figure=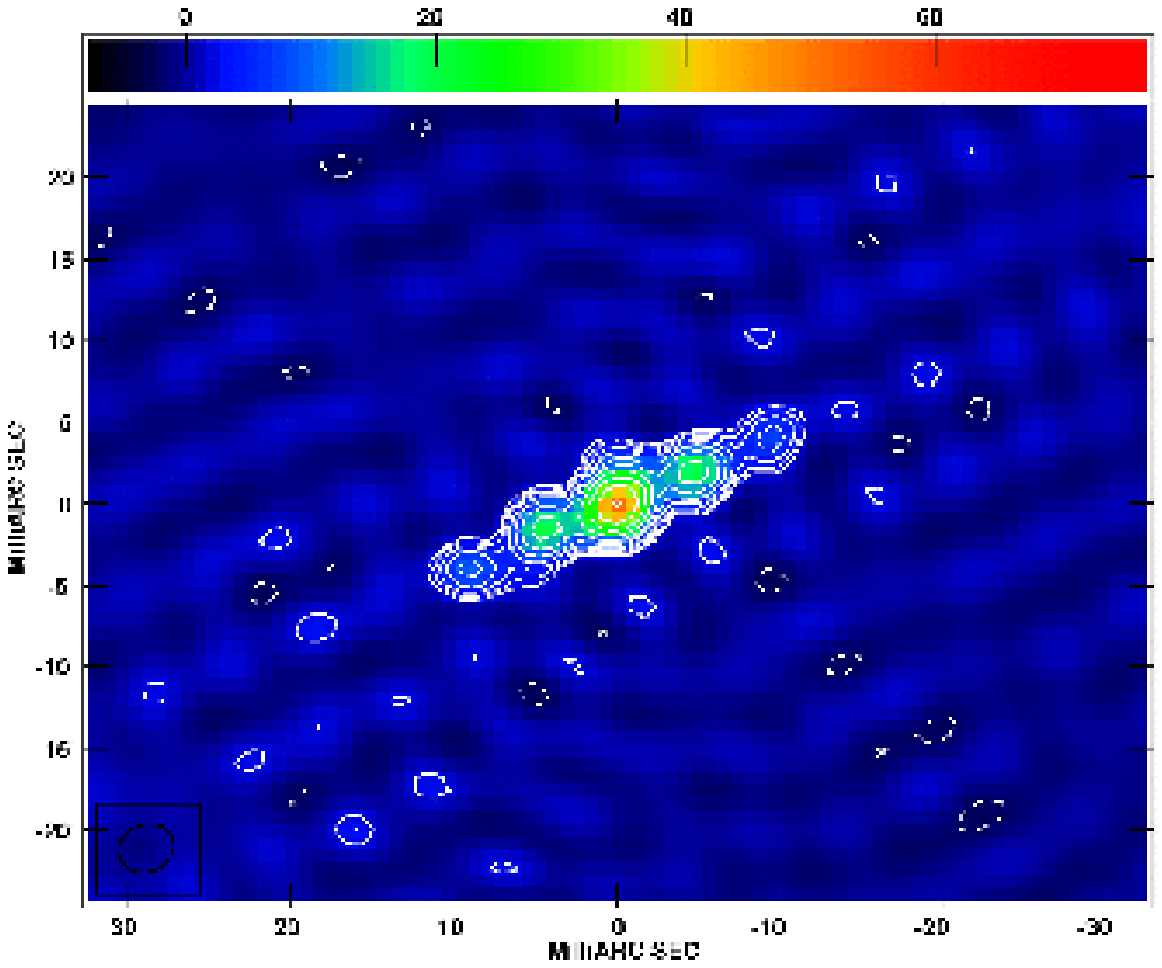,width=0.45\textwidth}}}
\caption{LBA images of Cir X-1 on 2010 July 28.  Contours for both images are at $\pm3\times(\sqrt{2})^n$ times the rms noise level, where $n=0,1,2,...$.  The colour bar shows the radio brightness in units of mJy\perbeam, with the scale being the same for both images.  Restoring beam size is $3.4\times2.9$\,mas$^2$ in P.A.\ $-69.1^{\circ}$.  The source is resolved into a continuous jet, which expands from an extent of 9.3\,mas to 19\,mas between the two halves of the observing run.}
\label{fig:lba_images}
\end{center}
\end{figure*}

Cir X-1 was resolved by the LBA into a symmetric, jet-like structure that evolved over the course of the observations (Fig.~\ref{fig:lba_images}).  Although our minimum and maximum baselines were 113 and 1702\,km, respectively, the failure at Tidbinbilla deprived us of intermediate baselines of order a few hundred kilometres.  This created a large hole in the {\it uv}-plane (Fig.~\ref{fig:uvcov}), such that we were only sensitive to structures on scales of 4--9 and 67--250\,mas.  Our single short baseline (between ATCA and Mopra) was consistent with a point source whose flux density gradually decayed from 210 to 80\,mJy\perbeam.

\begin{figure}
\centering
\includegraphics[height=0.95\columnwidth,angle=270]{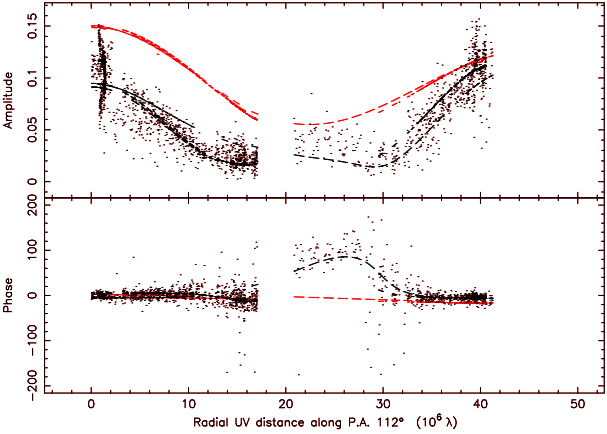}
\caption{Visibility amplitude and phase as a function of projected {\it uv}-distance along the position angle of the jet axis.  Plotted points show the data from the second half of the observation.  Red and black lines show the clean component models for the first and second halves, respectively.  The total amplitude decreases between the first and second halves, and the first minimum shifts to shorter projected baselines, implying that the overall structure has expanded. The relatively flat phases as a function of projected baseline length show the symmetry of the jets about the peak.}
\label{fig:projected}
\end{figure}

The changing flux density over the 14.3\,h observation violates one of the fundamental assumptions of aperture synthesis.  In the absence of morphological variations, this would create amplitude errors in the image plane.  However, since our {\it uv}-coverage is already so sparse, we cannot compensate by only imaging short snapshots of data in which the amplitudes are relatively stable.  Tests indicated that the data could not be binned more finely than into two separate halves without severely compromising the imaging fidelity.  A comparison of the data from the two halves of the observation (Fig.~\ref{fig:lba_images}) shows a fading of the core jets, and the appearance of two new outer components in the second half.  If these new components correspond to outward motion of the core jets seen in Fig.~\ref{fig:lba_images}(a), the combined proper motion of the two components is $\sim35$\,mas\,d$^{-1}$, significantly different from the 400\,mas\,d$^{-1}$ inferred by \citet{Fen04} from the time delay between core and lobe flaring events.  The expansion of the jets corresponds to the appearance of structures on larger scales, reflected in the shift to shorter spacings of the first minimum in the amplitudes of the visibility data projected along the jet axis (Fig.~\ref{fig:projected}, top).

To evaluate the effects of the changing source flux density and morphology combined with our relatively sparse {\it uv}-coverage, we performed some simulations within {\sc aips}.  We subtracted the clean components from our best image (Fig.~\ref{fig:lba_images}) in the {\it uv}-plane, and then added various source models back in to the {\it uv}-data.  A decaying point source at the phase centre as seen on the ATCA-Mopra baseline could not reproduce the symmetric extended structure along the observed position angle.  We therefore conclude that the observed structure is not an artifact caused by amplitude errors combined with limited {\it uv}-coverage.  Adding a second, non-varying component along the position angle of our observed jet showed that we could not reproduce the observed symmetric structure with an asymmetric, one-sided jet model.  Adding instead a single, ballistically moving jet component created a one-sided, smooth jet, with a reduced surface brightness owing to the smoothing of the emission along the component trajectory, and a slightly curved jet locus \citep[as also found by][]{Tin95}.  Artifacts also appeared as stripes parallel to the jet trajectory, reminiscent of the noise structure in Fig.~\ref{fig:lba_images}(a).  From these simulations, we conclude that while the relatively continuous appearance of the jets could be created by discrete, moving components, the symmetric profile we observe is real, as it cannot be reproduced using any one-sided source model.  The reality of the symmetric structure in the first half of the observations is also demonstrated by the flat and zeroed phases of the {\it uv}-data when projected along the jet axis (Fig.~\ref{fig:projected}).  Some slight asymmetry on scales of 7--10\,mas is visible in the second half.

Finally, we tested whether the extra components seen in Fig~\ref{fig:lba_images}(b) could have been detected with the {\it uv}-coverage available in the first half of the observations.  Substituting in the {\it uv}-plane the clean components of Fig.~\ref{fig:lba_images}(b) for those of Fig.~\ref{fig:lba_images}(a), we found that the extra emitting components, if present in the first part of the observations, would have been clearly visible in the images.  Either the jets have moved outwards between the two halves of the observation, or new components have become visible, via energisation of the electrons or amplification of the magnetic field.

The observed jets are elongated along a position angle 112\degree\ E of N, in good agreement with the position angle of the arcsecond-scale jets detected in both the radio and X-ray bands. \citet{Fen98} derived a position angle of $110\pm10$\degree\ for the extended arcsecond-scale radio emission, although \citet{Tud08} found this position angle to vary between epochs, with a mean of $129\pm13$\degree, although there was no unequivocal evidence for precession.  The deepest X-ray images \citep{Sel10} show unresolved emission at a position angle 140\degree\ E of N, and extended diffuse emission from 95--155\degree\ at a distance of 20--50\arcsecond\ from the core.  From the slightly different orientations seen on different angular scales, \citet{Cal11} suggested either a varying degree of precession with time or bends in the jet.

We find no evidence for angular broadening of the radio emission, as reported by \citet{Phi07} at the lower frequency of 1.6\,GHz.  Their measured angular size of $60\pm15$\,mas, scaled by the expected $\lambda^{2.2}$ dependence corresponds to $1.6\pm0.4$\,mas at our observing frequency of 8.4\,GHz, well below our beam size of $3.4\times2.9$\,mas$^2$.

\section{Discussion}
\label{sec:discussion}

\subsection{Morphology}
\label{sec:morphology}

The symmetry of the jets is initially reminiscent of the compact jets observed in the hard state of the black hole system GRS\,1915+105 \citep*{Dha00}, although we note several caveats to this interpretation.  The evolution of the source structure over the duration of the observing run (Fig.~\ref{fig:lba_images}) implies that this is not a persistent, compact jet.  Second, while we have no spectral information for this particular flare, previous multi-frequency observations of Cir X-1 have shown that the spectra evolve from optically thick to optically thin over the course of an outburst \citep{Fen05,Cal10}, and are usually optically thin by orbital phase 0.05 (G.D.~Nicolson, priv.\ comm.).  Thus it is likely that the observed emission is caused by the outward motion of discrete, expanding, optically-thin ejecta, as seen during outbursts of black hole X-ray binaries \citep[e.g.][]{Mir94,Tin95}.

Since the observations were not phase-referenced, we have no absolute positional information.  Also, the astrometric parameters of Cir X-1 are unknown, so the predicted location of the binary system cannot be accurately determined.  It is therefore uncertain whether the peak brightness corresponds to the location of the binary system, or rather to the surface of optical depth unity in the approaching jet.  The latter possibility would require a symmetric profile for the approaching jet, and for the receding jet to be hidden either by Doppler de-boosting, or by free-free absorption within a few hundred au of the central binary.  However, the apparently symmetric nature of the expansion seen in Fig.~\ref{fig:lba_images} tends to argue against this scenario.  The most plausible explanation for the observed symmetry is therefore an intrinsically slow jet such that the Doppler factors of approaching and receding jets are comparable.  In that case, the jets are only mildly relativistic and inclined close to the plane of the sky.

\subsection{Jet speed}

An ultrarelativistic flow with a proper motion of 400\,mas\,d$^{-1}$ would be smeared over 68 beams over the course of our 14.3\,h observation.  Unless the ejecta were intrinsically extremely bright ($>112$\,mJy), this would reduce the surface brightness below our $3\sigma$ detection threshold.  However, in the scenario of \citet{Fen04}, the postulated ultrarelativistic flow was unseen, but inferred from the brightening of radio-emitting components downstream in the jets as they were energized by this faster outflow.  A similar model was used to explain the brightening of radio-emitting lobes moving away from the neutron star X-ray binary Sco X-1 \citep*{Fom01}.  While the lobes in Sco X-1 moved at speeds of 0.32--0.57$c$, energy appeared to move from the core to the lobes at $>0.95c$.  A similar phenomenon has also been seen in the jets of active galactic nuclei \citep[e.g.][]{Tin98}.  Thus we might not expect to directly detect radio emission from any ultrarelativistic flow.

\begin{figure}
\centering
\includegraphics[width=0.95\columnwidth]{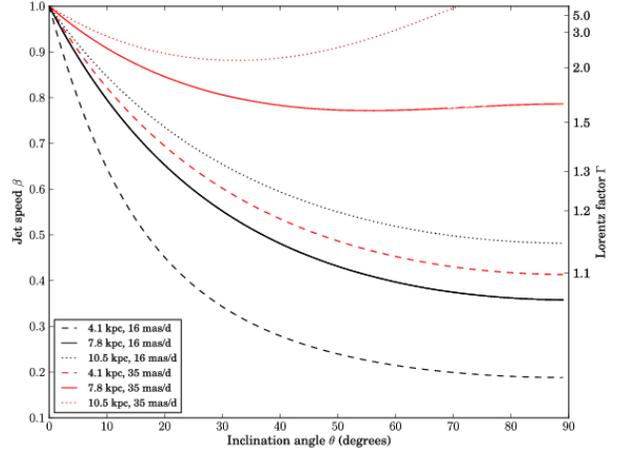}
\caption{Intrinsic jet speed as a function of inclination angle, given the proper motion constraints of $\mu_{\rm app}+\mu_{\rm rec}\approx16$\,mas\,d$^{-1}$ (black curves) and 35\,mas\,d$^{-1}$ (red curves).  Curves are plotted for the assumed distance, as well as for the lower and upper bounds of the commonly accepted distance range for the source.  Our proper motion constraints are both compatible with a mildly relativistic jet velocity at any moderate inclination angle, unless the distance is at the upper end of the estimated range.}
\label{fig:beta}
\end{figure}

Since compact radio emission from Cir X-1 is only present at and shortly after periastron \citep{Moi11}, we can assume that the jets were energized by the flare at orbital phase zero.  If an ultrarelativistic flow causes the brightening of the extra components in Fig.~\ref{fig:lba_images}(b), we can use the 1.2\,d delay between phase zero (from our ephemeris) and the time of the observations to infer its speed.  The observed size of $19$\,mas then gives a combined proper motion of approaching and receding jets of $\sim16$\,mas\,d$^{-1}$.  This value is even smaller than the combined proper motion of 35\,mas\,d$^{-1}$ calculated from the expansion seen in Fig.~\ref{fig:lba_images}, suggesting that the actual jet ejection might have come slightly later than phase zero.  Although these proper motion estimates do not allow us to uniquely determine the jet speed and inclination angle, for a given distance we can determine the jet speed as a function of inclination (Fig.~\ref{fig:beta}).  Unless either the jets are inclined extremely close to the line of sight, or the distance is at the upper end of the allowed range and the estimate of 35\,mas\,d$^{-1}$ is most representative, the intrinsic speed of the radio-emitting components is only mildly relativistic.

A flow travelling at 400\,mas\,d$^{-1}$ should by contrast have propagated out to 0.5\,arcsec by the end of our observations.  Imaging out to 0.8\,arcsec from the field centre showed no additional emission downstream in the jets, so any downstream knots energized by an ultrarelativistic flow must have been sufficiently diffuse to be resolved out.  Should an ultrarelativistic flow exist, the jets must be aligned within 5\degree\ of the line of sight \citep{Fen04}.  While reconciling this constraint with the symmetric morphology discussed in Section~\ref{sec:morphology} is difficult, the uncertainties in our interpretation are sufficient that we cannot definitively rule out this scenario.

\subsection{Opening angle}

Owing to our sparse {\it uv}-coverage coupled with intrinsic source variability, we cannot derive meaningful constraints on the opening angle from the visibility data.  In the image plane, the jets are not significantly resolved transverse to the jet axis.  From the length of the jets and the beam size of $3.4\times2.9$\,mas$^2$ in PA -69.1\degree, we constrain the opening angle of the jets to be $<20$\degree. This allows us to rule out poor collimation as an explanation for the half-opening angles of 35\degree\ found for the X-ray caps observed by \citet{Sel10}.  The remaining options put forward by the authors are precessing jets or a jet axis aligned very close to the line of sight.  The variation in the position angle of the extended radio emission and the symmetry of the milliarcsecond-scale structure (Section \ref{sec:results}) when compared with previous arcsecond-scale radio observations \citep{Cal11} would seem to support the precession scenario, although the standard deviation of the position angles (13\degree) is significantly smaller than the half-opening angle of the X-ray caps, and the time sampling is currently too sparse to determine a precession period.  Multiple VLBI measurements of the position angle of the jets would be required to test the precession scenario.

\section{Conclusions}

We have resolved the milliarcsecond-scale jets in Circinus X-1 for the first time.  The jet morphology is symmetric, extended on a scale of $\sim20$\,mas along a position angle 112\degree\ east of north, in good agreement with the position angle seen in ATCA observations of the arcsecond-scale jets.  We do not resolve the jets along the minor axis, implying an opening angle of $<20$\degree.  The source brightness decayed over the course of the observing run, and there is evidence for outward motion of the components between the first and second halves of the observation, implying a separation speed of $\sim35$\,mas\,d$^{-1}$.  This is greater than the proper motion estimate derived assuming that the jets were launched at the onset of the radio flare at orbital phase zero, implying a possible delay in the launching of the jets following periastron.  While the ultra-relativistic flow inferred by \citet{Fen04} cannot be definitively ruled out, the symmetry of the observed jets is difficult to reconcile with this scenario.

\section*{Acknowledgments}

The International Centre for Radio Astronomy Research is a joint venture between Curtin University and the University of Western Australia, funded by the state government of Western Australia and the joint venture partners.  The Australian Long Baseline Array is part of the Australia Telescope which is funded by the Commonwealth of Australia for operation as a National Facility managed by CSIRO.  SJT is a Western Australian Premier's Fellow.  AM was supported by a Curtin University international postgraduate scholarship and was a CSIRO-co-supervised PhD student.  This work made use of NASA's Astrophysics Data System.

\label{lastpage}
\bibliographystyle{mn2e}

\end{document}